\documentclass[english,floatfix,twocolumn,showkeys,superscriptaddress,nofootinbib,aps,prd]{revtex4}
\usepackage[latin1]{inputenc}
\usepackage[T1]{fontenc}
\usepackage{lmodern}
\usepackage{amsmath}
\usepackage{amssymb}
\usepackage{graphicx}
\usepackage{esint}
\usepackage{longtable}
\usepackage{dcolumn}
\usepackage{babel} 
\usepackage{csquotes} 
\usepackage{color} 

\begin{document}

\title{Bumblebee field as a source of cosmological anisotropies}

\author{R. V. Maluf}
\email{r.v.maluf@fisica.ufc.br}
\affiliation{Universidade Federal do Ceará (UFC), Departamento de Física,\\ Campus do Pici, Fortaleza - CE, C.P. 6030, 60455-760 - Brazil}
\author{Juliano C. S. Neves}
\email{juliano.neves@unifal-mg.edu.br}
\affiliation{Instituto de Ciência e Tecnologia, Universidade Federal de Alfenas, \\ Rodovia José Aurélio Vilela,
11999, CEP 37715-400 Poços de Caldas, MG, Brazil}

\begin{abstract}
In this work, a bumblebee field is adopted in order to generate cosmological anisotropies. For that purpose, we
assume a Bianchi I cosmology, as the background geometry, and a bumblebee field coupled to it. 
Bumblebee models are examples of a mechanism for the Lorentz symmetry violation by
assuming a nonzero vacuum expectation value for the bumblebee field. When coupled to 
the Bianchi I geometry, which is not in agreement with a cosmological principle, the bumblebee field
plays the role of a source of anisotropies and produces a preferred axis. Thus, a fraction of the cosmic anisotropies 
would come from the Lorentz symmetry violation.
In the last part of the article, we try to assume an upper bound on the bumblebee field using 
the quadrupole and octopole moments of the cosmic microwave background radiation.  
\end{abstract}

\keywords{Lorentz Symmetry Breaking, Cosmology, Cosmic Microwave Background, Cosmological Principle }

\maketitle

\section{Introduction}

The standard model in cosmology, namely the $\Lambda$CDM model or the big bang model, is based on 
the cosmological principle, 
alongside Einstein's field equations and the matter-energy description by means of perfect fluids.\footnote{For
many people, inflation is also considered an ingredient in the standard model.} Here, in a
gravitational model beyond general relativity, the bumblebee gravity, we will relax the cosmological
principle assumption and adopt other gravitational field equations, i.e., we will adopt modified
field equations instead of Einstein's.

As we said, the cosmological principle is a cornerstone in today's cosmology. We can define it by saying
the cosmological principle is the \textit{belief} according to which any observer or freely falling observer (like us),
 no matter where he/she is, describes
the same universe, the same physics laws or, philosophically speaking, the same phenomena \cite{Weinberg,Neves}. 
That is to say, 
freely falling observers would describe the same world, our universe would be homogeneous and isotropic
 for those observers. However, recent precise observations \cite{Planck1,Planck2} suggest
that the cosmological principle is no longer absolutely reliable: cosmic microwave background (CMB) anisotropies 
 would indicate different temperatures for different space directions.

As is well known, anisotropies in the CMB do not mean always a real problem for the standard model. There are
a lot of sources for that anisotropies, whether those ones in the present-day or  in the early universe. 
According to textbooks \cite{Weinberg}, the motion of the solar system and the Sunyaev-Zel'dovich \cite{Sunyaev} 
effect are examples
of  causes of anisotropies in the CMB in the recent universe. On the other hand, there are sources of anisotropies
 whose origin is in the early universe, like the Sachs-Wolfe effect \cite{Sachs} or even the inflationary period.
However, supposedly, there are still anisotropies  whose origin are not fully known, that is to say,
there is an intense debate on the reality or not (as statistical artefacts or not) of those anisotropies and 
on how isotropic our universe is \cite{Saadeh}.
Among those anisotropies or anomalies, we have the alignment of the quadrupole and octopole moments \cite{alignment},
the axis of evil \cite{axis_of_evil1,axis_of_evil2,axis_of_evil3}, and the cold spot \cite{cold_spot}.
Proposals have been built in order to solve those anomalies in the CMB, like inhomogeneous
spacetimes, local voids and even by using the inflationary mechanism (see Refs. \cite{Land:2005cg,Inoue:2006rd,Donoghue:2007ze,Koivisto:2008xf}).
Following Russell \textit{et al.} \cite{Russell}, we adopt the Bianchi I geometry to think of those anisotropies and
quantify them by means of the model-independent approach of Maartens \textit{et al.} \cite{Maartens}. 
However, in this article, we propose a source or origin for a part of the anisotropies in the CMB radiation by
using a Lorentz symmetry breaking mechanism. 
 
Then we suggest a source for CMB anisotropies whose origin is due to a field, the bumblebee field, which
 has broken the Lorentz symmetry or invariance since the decoupling 
 between matter and radiation. Therefore, in order to describe small cosmic anisotropies, 
 we will use modified gravitational field equations and
a special field coupled to the geometry. As will see, the modified field equations come from
the bumblebee gravity. On the other hand, as we said, the geometry is the Bianchi I metric, which provides
three independent directional Hubble parameters, one for each space direction. 

The Bianchi I geometry has been studied in several contexts beyond general relativity \cite{Alexeyev:2000eb,Saha:2006iu,Rao:2008zzd,Rikhvitsky:2011js,Carloni:2013hna}. 
Such a geometry generalizes the 
flat Friedmann-Lemaître-Robertson-Walker (FLRW)
adopted in the standard model. The Bianchi I spacetime defines three different scale factors, one for 
each spatial direction. When all those three are identical ones, the FLRW metric is restored. 
 In the general relativity context, the difference among each directional Hubble parameter is 
 due to  integration constants that come out 
from each differential equation related to the three directional Hubble parameters. But in the bumblebee gravity, 
as we propose here, such a difference in each directional Hubble parameter
appears because of the bumblebee field, when such a quantum field assumes a nonzero vacuum expectation value (VEV).
The idea is that the bumblebee field in the VEV breaks spontaneously the Lorentz symmetry in the
decoupling period, when matter and radiation decouple. Thus, in the model
developed here, a part of the cosmic anisotropies is due to the Lorentz violation from the decoupling period.
But before the present time or from the decoupling to the present time, our model is a small deviation from
the FLRW universe as it is convenient in order to agree with the data.
Therefore, the key feature of the model present here is the Lorentz symmetry or invariance violation, which
would produce cosmological anisotropies.   

The possibility of the Lorentz invariance violation in the gravitational context was initially discussed
 by Kostelecký   \cite{Kostelecky2004} in 2004. 
 In that seminal work, a no-go theorem was presented stating that explicit Lorentz-violation theories,
  containing fixed and nondynamical background fields, do not preserve geometric constraints and 
  fundamental conservation laws in general relativity. One way to get around these difficulties is 
  to consider some spontaneous Lorentz symmetry breaking mechanism, something which works like the Higgs
   mechanism, by adding a potential term to the Lagrange density able to generate a non-trivial vacuum state, 
   where either vector or tensorial fields assume nonzero VEVs.

A mechanism like the bumblebee  has been applied in black
hole physics and cosmology. In black hole physics, a Schwarzschild-like \cite{Casana,Gullu} and both a 
Schwarzschild-de Sitter-like and a Schwarzschild-anti-de Sitter-like
  spacetimes \cite{Maluf_Neves} were built. In Ref. \cite{Bertolami} other types of black holes were obtained as well.
  In cosmology, consequences of the bumblebee field in the
FLRW geometry were studied \cite{Capelo,ONeal}, and works on the Gödel universe were presented in Refs. \cite{Petrov,Jesus}
in the same context of Lorentz violation. As far as we know, using the Bianchi I metric in this context is the
first attempt in that direction.

This article is structured as follows: in Sec 2 we present both the modified gravitational field equations in 
bumblebee gravity and the Bianchi I geometry in that context, solving then the field equations for a universe 
made up of radiation, matter and the bumblebee field. In Sec. 3 we quantified anisotropies and use 
the CMB data from Planck Collaboration in order to constrain the
 bumblebee field (and the coupling constant) by means of the  CMB multipoles. The final remarks are given in Sec. 4. 
 
\section{Bianchi I cosmology in bumblebee gravity}
In this section, the bumblebee model is introduced. The modified field equations in that context will be solved
using the Bianchi I metric as our \textit{Ansatz}. Friedmann-like equations will be obtained and conceived of as
small deviations from the standard model equations. Following Russell \textit{et al.} \cite{Russell}, 
the main idea here is to describe the cosmological expansion from the decoupling period until the present time.

\subsection{The gravitational field and the bumblebee field equations}
Bumblebee models are vector or tensor theories that include some mechanism for describing spontaneous 
breaking of the Lorentz and diffeomorphism symmetries within a gravitational context. 
These models have a potential term $V$ which leads to nontrivial VEVs 
for the fields configurations, affecting, for example, the dynamics of other fields coupled to the 
bumblebee field,  preserving geometric structures and conservation laws that are compatible with 
a usual pseudo-Riemannian manifold adopted in general relativity \cite{Kostelecky2004,Bluhm2005}.

The simplest model involving a single vector field $B_{\mu}$ (the bumblebee field) 
coupled to gravity in a torsion-free spacetime is described by the action
\begin{eqnarray}
 S_{B}&=&\int d^{4}x\sqrt{-g} \bigg [  \frac{1}{2\kappa} R+\frac{\xi}{2\kappa}B^{\mu}B^{\nu}R_{\mu\nu} \nonumber \\ 
 &&-\frac{1}{4} B_{\mu\nu}B^{\mu\nu}-V(B^{\mu}B_{\mu}\pm b^{2})+\mathcal{L}_{M} \bigg ], 
 \label{S1} 
\end{eqnarray}
 where $\kappa=8\pi G/c^4$ is the gravitational coupling constant, and $\xi$ is a coupling constant, 
 accounting for the nonminimum interaction 
 between the bumblebee field and the Ricci tensor or geometry (with mass 
 dimension $[\xi]=M^{-2}$ in natural units) \cite{Bailey2006,linearized}. Other important ingredients are
   $B_{\mu\nu}\equiv\partial_{\mu}B_\nu-\partial_{\nu}B_{\mu}$, or the bumblebee field strength, 
  and the Lagrange density $\mathcal{L}_{M}$, which describes the matter-energy content, 
  something very necessary in a cosmological model. 
  
As we said, the potential $V(X)$ in the action (\ref{S1}) is responsible for triggering a nonzero VEV
 for the bumblebee field and the metric, i.e.,
\begin{equation}
B_{\mu}\rightarrow \left\langle B_{\mu}\right\rangle = b_{\mu},\ \ \ \ \ g_{\mu\nu} \rightarrow \left\langle g_{\mu\nu},\right\rangle,\label{VacuumSolution}
\end{equation}
 thereby it  breaks spontaneously both  the Lorentz and the diffeomorphism symmetry. 
 For a smooth potential $V$ of $X$, the 
 vacuum condition $X = b_{\mu}\left\langle g^{\mu\nu}\right\rangle b_{\nu}\pm b^{2}=0$ implies
  that the potential and its derivative satisfy $V = V'=0$, where $\left\langle g^{\mu\nu}\right\rangle$ is the 
  VEV of the inverse metric. It is worth noticing that the quantity $b^{2}$ 
is a positive real number, and the $\pm$ sign implies that $b_{\mu}$ is timelike or spacelike, respectively.  
  
\textit{A priori}, field excitations around the vacuum solutions (\ref{VacuumSolution}) can happen, 
leading then to the emergence of massless Nambu-Goldstone modes and 
massive modes \cite{Bluhm2005,Bluhm2008}. Undoubtedly, establishing the phenomenological 
roles of these modes in the cosmological context is an interesting issue, but it is beyond the scope of this work.  
For the present purposes, we will assume that the field excitations are turned off and that both the bumblebee 
field and the metric are frozen at their VEVs as
\begin{equation}
B_{\mu}=b_{\mu},\ \ \ \ \ g_{\mu\nu} =\left\langle g_{\mu\nu}\right\rangle,
\end{equation}where, in general, $b_{\mu}$ and $\left\langle g_{\mu\nu}\right\rangle$ are functions of the spacetime position.
  
The sought-after gravitational field equations in the bumblebee gravity can be directly obtained  by varying the 
action (\ref{S1}) with respect to
the metric tensor $g_{\mu\nu}$, while keeping the bumblebee field $B_{\mu}$ totally fixed. That is, with
that procedure we have the following modified gravitational field equations: 
\begin{align}
 G_{\mu\nu}  = & \kappa \left( T^{B}_{\mu\nu}+ T^{M}_{\mu\nu} \right) \nonumber \\
	= &  \kappa\left[2V'B_{\mu}B_{\nu} +B_{\mu}^{\ \alpha}B_{\nu\alpha}-\left(V+ \frac{1}{4}B_{\alpha\beta}B^{\alpha\beta}\right)g_{\mu\nu} \right] \nonumber\\
		& +\xi\left[\frac{1}{2}B^{\alpha}B^{\beta}R_{\alpha\beta}g_{\mu\nu}-B_{\mu}B^{\alpha}R_{\alpha\nu}-B_{\nu}B^{\alpha}R_{\alpha\mu}\right.\nonumber\\
		& +\frac{1}{2}\nabla_{\alpha}\nabla_{\mu}\left(B^{\alpha}B_{\nu}\right)+\frac{1}{2}\nabla_{\alpha}\nabla_{\nu}\left(B^{\alpha}B_{\mu}\right) \nonumber\\
	& \left.-\frac{1}{2}\nabla^{2}\left(B_{\mu}B_{\nu}\right)-\frac{1}{2}
g_{\mu\nu}\nabla_{\alpha}\nabla_{\beta}\left(B^{\alpha}B^{\beta}\right)\right] +\kappa T^{M}_{\mu\nu}.
\label{modified}
\end{align}
$G_{\mu\nu}$ is the Einstein tensor, the operator $'$ means derivative with respect to the potential argument, and
 $T^{B}_{\mu\nu}$ and $T^{M}_{\mu\nu}$ are, respectively, the energy-momentum tensor of the bumblebee field and of the matter field.
 
In order to solve the very nonlinear Eq. (\ref{modified}), it is necessary to choose a metric \textit{Ansatz}---the 
Bianchi I metric in this article---and a bumblebee field. The explicit form of the potential $V$ is irrelevant 
because we are going to work in the VEV of the field, thus it preserves the vacuum condition $V=V'=0$.
 
Lastly, the action (\ref{S1}) also delivers an equation of motion for the bumblebee field. By varying that action, 
in this time with respect to the bumblebee field, we have
  \begin{equation}
 \nabla_{\mu}B^{\mu\nu}=2\left( V'B^{\nu}-\frac{\xi}{2\kappa}B_\mu R^{\mu\nu}  \right),
 \label{B_eq}
 \end{equation}
as the equation of motion for the field $B_{\mu}$. Besides, for the sake of simplicity, we suppose here that the matter sector does not couple with the bumblebee field.

\subsection{The Friedmann-like equations}
The Bianchi I model or cosmology, which generalizes the flat FLRW geometry, 
is given by the metric or line element
\begin{equation}
ds^2 = -c^2 dt^2 +a_{1}(t)^2 dx_{1}^2+a_{2}(t)^2 dx_{2}^2+a_{3}(t)^2 dx_{3}^2,
\label{Metric}
\end{equation}
in the $(t,x_1,x_2,x_3)$ coordinates. Here, $a_1(t),a_2(t)$, and $a_3(t)$ are directional scale factors, they
indicate different expansion rates for each spatial direction. Therefore, in such a spacetime, the Hubble parameter is defined in each spatial direction, i.e.,
\begin{equation}
H_{i}=\frac{\dot{a_{i}}(t)}{a_{i}(t)}, \\\ \text{with \textit{i}=1,2,3},
\end{equation}
where dot means derivative with respect to the temporal coordinate $t$ or the cosmic time. From different values
for each directional Hubble parameter, one has an anisotropic cosmology. 

For the matter-energy content, we assume the perfect fluid description in this article, that is to say, 
the energy-momentum tensor
for matter fields is described and given by the well-known expression:
\begin{equation}
T_{\mu\nu}^M = \left[\rho_{M}(t) + \frac{p_{M}(t)}{c^2}  \right]u_{\mu}u_{\nu}+p_{M}(t)g_{\mu\nu},
\label{PF}
\end{equation}
with $\rho_M (t)$ and $p_M(t)$ playing the role of the  density and pressure, respectively, of the matter-energy
 content (excluding the bumblebee field). The four-vector $u^{\mu}$ is the four-velocity of the fluid, and, in particular
 for the metric signature adopted here,  $u_{\mu}u^{\mu}=-c^2$. It is worth emphasizing that
 the perfect fluid is isotropic, and anisotropy comes from the bumblebee field.
 Contrary to works in bumblebee gravity in which
black holes in vacuum spacetimes were studied, in a \textit{realistic} cosmological context matter fields do matter.  

Our goal is to choose a particular bumblebee field that plays the role of a source of cosmic anisotropies.
Without loss of generality, a suitable choice for the bumblebee field, when it assumes the VEV, is written as
\begin{equation}
    B_\mu=b_\mu=(0,b(t),0,0),
    \label{B}
\end{equation}
in which the field component in the $x_1$ direction is just $t$-dependent.  As we said, the VEV 
for the bumblebee field is nonzero and its norm is constant, that is,  $b_\mu b^\mu=b^2$. Thus, from the Bianchi I metric (\ref{Metric}),
we have
\begin{equation}
b(t) = b a_1(t).
\end{equation}
Besides the condition $V(b_\mu b^{\mu}-b^2)=0$, for the sake of simplicity, we adopt $V'=0$ as well, excluding
contributions from the potential derivative. For Lagrange-multiplier potentials, 
as discussed in Ref. \cite{Maluf_Neves}, $V'$ can contribute to
the equations of motion. But, as we will see, even with $V'=0$ one has the bumblebee field
playing the role of a source for anisotropies.

With the choice indicated in Eq. (\ref{B}), the metric (\ref{Metric}) provides an equation from the 
bumblebee equation of motion (\ref{B_eq}).
It is worth noting that the field strength $B_{\mu\nu}$ is not identically zero according to the bumblebee field (\ref{B}). 
There are two nonzero components, namely $B_{01}=-B_{10}=\dot{b}(t)$. Therefore, by using (\ref{B_eq})
and the spacelike bumblebee field (\ref{B}), one has
\begin{equation}
\dot{H_1}+3HH_1-\frac{\kappa H_1^2}{\left(\kappa-\xi \right)}=0,
\label{Bumblebee_field}
\end{equation}
which it will be useful for the next calculations. The parameter $H$ stands for mean of the Hubble parameter, i.e.,    
\begin{equation}
H=\frac{1}{3}\left( H_1+H_2+H_3\right).
\end{equation}
Related to $H$, another useful quantity is the volume factor $\mathcal{V}$, which is written as a function of either the scalar
factor or the mean of Hubble parameter, i.e.,
\begin{equation}
\mathcal{V}=a_1 a_2  a_3  \hspace{0.5cm} \mbox{or} \hspace{0.5cm} \frac{1}{3}\frac{\dot{\mathcal{V}}}{\mathcal{V}}=H.
\label{volume}
\end{equation}

From the specific form for the bumblebee field $B_{\mu}$, given by Eq. (\ref{B}), and the Bianchi I metric (\ref{Metric}), 
the modified field equations give us the following energy-momentum tensor
 for the bumblebee field, namely
 \begin{equation}
 \left(T^B \right) _{\ \nu}^{\mu}= \left(\begin{array}{cccc}
-\rho_B c^2\\
 &p_1 \\
 &  & p_2 \\
 &  &  & p_3
\end{array}\right),
\label{Energy-momentum}
\end{equation}  
 with
\begin{equation}
\rho_B c^2 = p_2= p_3= \frac{\ell H_1^2}{2\xi c^2} \hspace{0.5cm} \mbox{and} \hspace{0.5cm} p_1=-\frac{ \left(\kappa+\xi\right) }{ \left(\kappa-\xi\right)}\frac{\ell H_1^2}{2\xi c^2},
\label{Components}
\end{equation}
where $\ell = \xi b^2$, the parameter that accounts for Lorentz-violation effects,  is commonly called Lorentz-violating parameter,  and $\rho_B$ and $p's$ are
the density and pressure, respectively, of the bumblebee field (note that the simplified result for $p_1$ is obtained by 
using the bumblebee equation (\ref{Bumblebee_field})). 
 As we can see, the bumblebee  energy-momentum tensor exhibits its anisotropic feature due to fact that
$p_1 \neq p_2 = p_3$. The bumblebee field modifies the $x^1$ direction. This will get even more evident from the
directional Hubble parameters. Therefore, even adopting a perfect fluid description for the matter-energy content,
anisotropies would arise from the bumblebee field as indicated in the energy-momentum tensor (\ref{Energy-momentum}).

With the gravitational field equations (\ref{modified}), the Bianchi I geometry (\ref{Metric}), and 
the bumblebee field equation (\ref{Bumblebee_field}), we obtained the Friedmann-like
equations in the bumblebee gravity adopted here. Then the $\mu=\nu=0$ component of the modified gravitational field
equations leads to
\begin{equation}
H_1H_2 + H_1H_3 + H_2H_3 =8\pi G \left(  \rho_M  +\rho_B  \right),
\label{G00}
\end{equation}
and the spatial  components read
\begin{align}
 \dot{H_2} + \dot{H_3} +H_2^2+H_3^2+H_2H_3  = & -\frac{8\pi G}{c^2} \left( p_M - p_1 \right),   \label{G11} \\
 \dot{H_1} + \dot{H_3} +H_1^2+H_3^2+H_1H_3  = &  -\frac{8\pi G}{c^2} \left(p_M  + p_2 \right),   \label{G22} \\
 \dot{H_1} + \dot{H_2} +H_1^2+H_2^2+H_1H_2  = &  - \frac{8\pi G}{c^2}\left(p_M + p_3 \right). \label{G33}
\end{align}
As we clearly see, the corresponding equation for the $\mu=\nu=1$ component, that is Eq. (\ref{G11}),
 is different from
other directional components, Eqs. (\ref{G22}) and (\ref{G33}),  due to the bumblebee field. 
By making $\ell=0$ and, consequently, $\rho_B=p_1=p_2=p_3=0$, one has the usual components of the Friedmann-like equations for the Bianchi I 
cosmology in the Einsteinian context. 

The matter-energy conservation provides another important equation. 
Using $\nabla _{\nu}(T^{\mu\nu}_B+T^{\mu\nu}_M)=0$ and the condition (\ref{Bumblebee_field}), 
we have the following relation for the energy-momentum tensor components of the matter fields:
\begin{equation}
\dot{\rho}_M=-3H \left(\rho_M+\frac{p_M}{c^2}\right).
\label{Conservation}
\end{equation}
Therefore, as we mentioned, the bumblebee field does not interact with the matter fields in this case. 

Adding up the spatial components of the modified gravitational field equations, Eqs. (\ref{G11})-(\ref{G33}),
with the aid of both the temporal component (\ref{G00})  and the useful relation
\begin{equation}
\sum_{i=1}^{3}H_i^2= 9H^2 - 2\left( H_1H_2 + H_1H_3 + H_2H_3\right),
\end{equation}
we obtain the Friedmann-like equation for the mean Hubble parameter:
\begin{equation}
\dot{H}+3H^2=  4\pi G \left[\rho_M + \frac{1}{3}\rho_B - \frac{1}{c^2} \left( p_M+\frac{1}{3}p_1 \right) \right].
\label{H}
\end{equation}

From Eq. (\ref{G00}), we calculate all three equations for the  directional Hubble
parameters with the aid of  Eqs. (\ref{G11})-(\ref{G33}) and the mean Hubble parameter (\ref{H}). Such equations are
written as
\begin{align}
 \dot{H_1}+3HH_1= &  \frac{\kappa c^2}{2} \left(\rho_M c^2 - p_M + \frac{2 \kappa p_1}{\left(\kappa+\xi \right)} \right) , \label{H1i} \\
 \dot{H_2}+3HH_2= & \frac{\kappa c^2}{2} \left(\rho_M c^2 - p_M - \frac{2 \kappa p_1}{\left(\kappa+\xi \right)} \right), \label{H2i} \\
 \dot{H_3}+3HH_3= & \frac{\kappa c^2}{2} \left(\rho_M c^2 - p_M - \frac{2 \kappa p_1}{\left(\kappa+\xi \right)} \right). \label{H3i}
\end{align}
It is possible to eliminate $H_1$ (expressed by the component $p_1$ of the energy-momentum tensor of 
the bumblebee field) of each directional Hubble parameter using the bumblebee
equation (\ref{Bumblebee_field}) and the Hubble parameter for the $x_1$ direction (\ref{H1i}), thus $H_1^2=\xi c^2 (\kappa-\xi)(\rho_M c^2-p_M)/(\ell \kappa+2\xi)$. As a result, fortunately, 
we are able to see each equation for the directional Hubble parameters as a small deviation from the FLRW metric.
That is to say, by doing that we have the sough-after equations for each directional Hubble parameter written as
a deviation from FLRW: 
\begin{align}
 \dot{H_1}+3HH_1= & 4\pi G \left(1-\delta \right) \left(\rho_M -\frac{p_M}{c^2} \right) , \label{H1} \\
 \dot{H_2}+3HH_2= & 4\pi G \left(1+\delta \right) \left(\rho_M -\frac{p_M}{c^2} \right), \label{H2} \\
 \dot{H_3}+3HH_3= & 4\pi G \left(1+\delta \right) \left(\rho_M -\frac{p_M}{c^2} \right), \label{H3}
\end{align}
with  the dimensionless positive constant $\delta$ given by
 \begin{equation}
\delta= \frac{\ell \kappa}{\ell \kappa+2\xi}\ll 1.
\label{delta}
 \end{equation}
  Hence, the bumblebee influence on the spacetime geometry is decoded in the parameter $\delta$.
  And, consequently, the equation for the mean Hubble parameter takes also a friendly form  given by
\begin{equation}
\dot{H}+3H^2=  4\pi G \left(1+\frac{\delta}{3} \right) \left(\rho_M -\frac{p_M}{c^2} \right).
\label{H_final}
\end{equation}

As we can see from Eqs. (\ref{H1})-(\ref{H3}), the equations for the components $H_2$ and $H_3$ are identical. 
On the other hand, for the
component $H_1$, Eq. (\ref{H1}) 
provides a different solution compared to Eqs. (\ref{H2})-(\ref{H3}). Thus, the difference among the Hubble parameter
components comes from the component in the $x_1$ direction, which could be seen as a preferred axis. 
Contrary to the Bianchi I cosmology 
in the Einsteinian context, anisotropies have the field $B_\mu$ as its source in the bumblebee gravity
adopted here. With
$\delta =0$, one has the Einsteinian context and identical differential equations for each directional
 Hubble parameter. Then, in that case, in the general relativity realm, as indicated in Ref. \cite{Russell}, 
  anisotropies come out of different integration constants  from each Hubble parameter differential equation.
 Here, we point out to the bumblebee vector field as a source (among others) of anisotropies. 
 
 As already mentioned in Introduction, the Lorentz violation is triggered by the 
 potential $V(b_\mu b^\mu -b^2)$ when the
 bumblebee field  assumes its VEV. We can observe the Lorentz symmetry violation looking at
 Eqs. (\ref{H1})-(\ref{H3}) and their differences, which come from the term that contains the Lorentz-violating parameter
 $\ell$. The Lorentz violation is translated into a preferred direction, the $x_1$ direction in our case, and
 the bumblebee field as its origin. Solutions of (\ref{H1})-(\ref{H3}) are written as
 \begin{equation}
 H_1(t)=\frac{1}{\mu (t)}\bigg[ K_1 + \int^{t}  \mu (t') 4\pi G \left(1-\delta \right) \bigg(\rho_M -\frac{p_M}{c^2} \bigg) dt' \bigg],
 \label{H1_solution}
 \end{equation}
 for the component in the $x_1$ direction, and 
 \begin{equation}
 H_j(t)=\frac{1}{\mu (t)}\bigg[ K_j + \int^{t} \mu (t') 4\pi G \left(1+ \delta \right) \bigg(\rho_M -\frac{p_M}{c^2} \bigg) dt' \bigg],
 \label{Hj_solution}
 \end{equation}
 for $j=2,3$ or the components in the $x_2$ and $x_3$ directions, respectively. 
 In these equations, $K_i$ with $i=1,2,3$ represent the mentioned integration constants. And the function $\mu (t)$ 
 is defined as
\begin{equation}
\mu (t)= \exp \left(\int^{t} 3H(s) ds\right).
\label{mu}
\end{equation} 
 
In order to have the bumblebee field as source or origin of the anisotropies of the model, we choose $K_1=K_2=K_3=0$
 in the equations above.  Then the geometry described by Eqs. (\ref{H1_solution})-(\ref{Hj_solution})
  presents the planar symmetry in the $x_2-x_3$ plane, and the anisotropic feature of that spacetime comes from $\delta$.
  For $\delta \ll 1$, one has a small deviation from isotropy, and as the integration constants $K_1$, $K_2$, and
  $K_3$ are set to zero, the geometry given by Eqs. (\ref{H1_solution})-(\ref{Hj_solution}) is a small deviation from 
  the FLRW  metric. It is worth pointing out that the factor $(1\pm \delta)$ can not be absorbed into the constant
  $G$, turning the solution (\ref{H1_solution})-(\ref{Hj_solution}) into the FLRW metric. As we have
  minus delta for the $H_1$ component and plus delta for the
  $H_2$ and $H_3$ components, the difference of  space directions is irremovable.
  
\subsection{Solving the Friedmann-like equations}
The basic idea here is to build a solution of (\ref{H1_solution}) and (\ref{Hj_solution}) 
for a universe made up of matter and radiation that evolves into a 
matter-dominated universe. Thus, following Russell \textit{et al.} \cite{Russell}, we do not take in
consideration dark energy. Here we assume that the bumblebee field and, consequently, the Lorentz violation
is triggered at the decoupling. Hence, tiny \textit{fingerprints} of the symmetry breaking would be in the CMB
translated into tiny anisotropies. Therefore, anisotropies from the Lorentz violation would be due to an event
whose beginning is found in the decoupling period, contrary 
to the dipole anisotropy and the Sunyaev-Zel'dovich effect that are present-day effects. 

The decoupling period is identified to an event right after the mater-radiation equality, given by the
 redshift $z_{eq} \simeq 3300$.  Then our spacetime evolves dynamically from a two-component universe (matter
 and radiation) to a matter-dominated one. The bumblebee field would be a third component
 in this model. However, as we interpret (\ref{H1})-(\ref{H3}) as a small deviation from the FLRW, 
 the entire contribution from the bumblebee field is decoded in the factor $\delta$.
 
 Having said that, radiation and matter are decoupled, then their equations of state can
 be written as
 \begin{equation}
 p_{r}=\omega_r \rho_{r}c^2=\frac{1}{3}\rho_r c^2 \hspace{0.5cm} \mbox{and} \hspace{0.5cm} p_m=\omega_m \rho_mc^2= 0, 
 \end{equation}
 with, as it is clear, $\omega_r=1/3$ and $\omega_m=0$. Using (\ref{H_final}), the dynamics of 
 the Bianchi I in the bumblebee  gravity can be evaluate from
\begin{equation}
\dot{H}_{rm}+3H_{rm}^2=  \frac{\kappa c^4}{2}\left(1+\frac{\delta}{3} \right)  \bigg(\frac{2}{3}\rho_r  + \rho_m  \bigg),
\label{Hrm}
\end{equation}  
 in the radiation-matter universe ($H_{rm}$ is the mean directional Hubble parameter for a  universe
 made up of radiation and matter). It is appropriate to write the densities in terms of their today's values. Following \cite{Russell},
 we are going to build the directional Hubble parameters in this context from the total volume of those periods 
 (radiation-matter content and matter-dominated universe). For that purpose, Eq. (\ref{Conservation}) provides
useful relations between both $\rho_M=\rho_r+\rho_m$ and the Hubble parameter (and consequently the volume factor) given by
\begin{equation}
\dot{\rho}_r=-4H_{rm}\rho_r \hspace{0.5cm} \mbox{and} \hspace{0.5cm} \dot{\rho}_m=-3H_{rm}\rho_m,
\end{equation}
which, after integration, lead to the normalized densities, i.e.,
\begin{align}
\rho_r = & \rho_{r,0} \left( \frac{\mathcal{V}_{rm,0}}{\mathcal{V}_{rm}}\right)^{\frac{4}{3}},\label{rho_r} \\
\rho_m= & \rho_{m,0} \left( \frac{\mathcal{V}_{rm,0}}{\mathcal{V}_{rm}}\right) \label{rho_m},
\end{align}
where the definition of $\mathcal{V}$, indicated in Eq. (\ref{volume}),
 was used. Here $\rho_{r,0}$ and $\rho_{m,0}$ are the present-day values 
for the radiation and matter densities, respectively,
 $\mathcal{V}_{rm}$ is the volume factor of a universe made up of both radiation and matter, 
 and $\mathcal{V}_{rm,0}$ 
 is the normalized present-day
 volume factor ($\mathcal{V}_{rm,0}=1$). 

As we have a flat FLRW or, at most, a small deviation from that spacetime in the today's 
observations \cite{Planck1,Planck2}, we assume that the total density parameter is given by
\begin{equation}
\Omega_T =\frac{\rho_T}{\rho_c} = 1 \hspace{0.5cm} \mbox{with} \hspace{0.5cm} \rho_c=\frac{3H_{0}^{2}}{8\pi G},
\label{Critical_density}
\end{equation}
in agreement with a spatially flat Bianchi I spacetime (\ref{Metric}), where $\rho_T$ and $\rho_c$ are
the total density and the critical density, respectively, and $H_0$ is the present-day (mean) Hubble factor.  
 It is worth mentioning that the normalized densities (\ref{rho_r})-(\ref{rho_m}) are very useful in order to 
 compare parameters to the recent data.
They are given by the density parameters and the present-day 
 value for the (mean) Hubble parameter, for $\rho_{r,0}$ and $\rho_{m,0}$, according to Eq. (\ref{Critical_density}),
  are written as
 \begin{equation}
 \rho_{r,0}=\frac{3H_0^2}{8\pi G}\Omega_{r,0} \hspace{0.5cm} \mbox{and} \hspace{0.5cm} \rho_{m,0}=\frac{3H_0^2}{8\pi G}\Omega_{m,0}.
 \end{equation}
The above relations are valid due to assumption  that the geometry
studied here is asymptotically FLRW as $t \rightarrow t_0$.

 Following Russell \textit{et al.} \cite{Russell}, firstly we are going to obtain the volume for a universe made up of matter and radiation and
 then the corresponding Hubble parameter or each directional Hubble parameter. From Eq. (\ref{Hrm}), it follows that
 \begin{equation}
 \ddot{\mathcal{V}}_{rm}- \left(1+\frac{\delta}{3} \right) \left(\frac{3H_0^2}{\mathcal{V}_{rm}^{1/3}}\Omega_{r,0}+\frac{9}{2}H_0^2\Omega_{m,0} \right)=0.
 \label{Vdotdot}
 \end{equation}
Multiplying (\ref{Vdotdot}) with $\dot{\mathcal{V}}_{rm}$ gives us 
\begin{equation}
\dot{\mathcal{V}}_{rm}^2-9H_0^2 \left(1+\frac{\delta}{3} \right) \left(\Omega_{r,0}\mathcal{V}_{rm}^{\frac{2}{3}} +\Omega_{m,0}\mathcal{V}_{rm} \right)=0,
\end{equation}
whose approximate solution is written in terms of the volume of a matter-dominated universe   plus
the volume of the period in which radiation and matter had equal densities, also known as radiation-matter 
equality, that is to say, the volume factor reads
\begin{equation}
\mathcal{V}_{rm} \simeq \frac{9}{4}\bigg(1+\frac{\delta}{3} \bigg) H_0^2 \Omega_{m,0}t^2 + 5\bigg( \frac{\Omega_{r,0}}{\Omega_{m,0}}\bigg)^3 = \mathcal{V}_m + 5\mathcal{V}_{rme},
\label{Vrm}
\end{equation}  
in which $\Omega_{r,0}/ \Omega_{m,0} \sim 10^{-4}$. It is easy to obtain the 
corresponding mean Hubble parameter from Eq. (\ref{Vrm}):
\begin{equation}
H_{rm} \simeq \frac{2}{3t}\frac{1}{\left(1+5\frac{\mathcal{V}_{rme}}{\mathcal{V}_m} \right)}.
\label{Hubble_rm}
\end{equation}
As we can see, for a matter-dominated universe, $\mathcal{V}_m \gg \mathcal{V}_{rme}$, thus one has the well-known
Hubble parameter for such a period, i.e., $H_{rm} \simeq H_m=\frac{2}{3t}$. 

\begin{figure}
\begin{centering}
\includegraphics[trim=0.55cm 0cm 2cm 0cm, clip=true,scale=0.44]{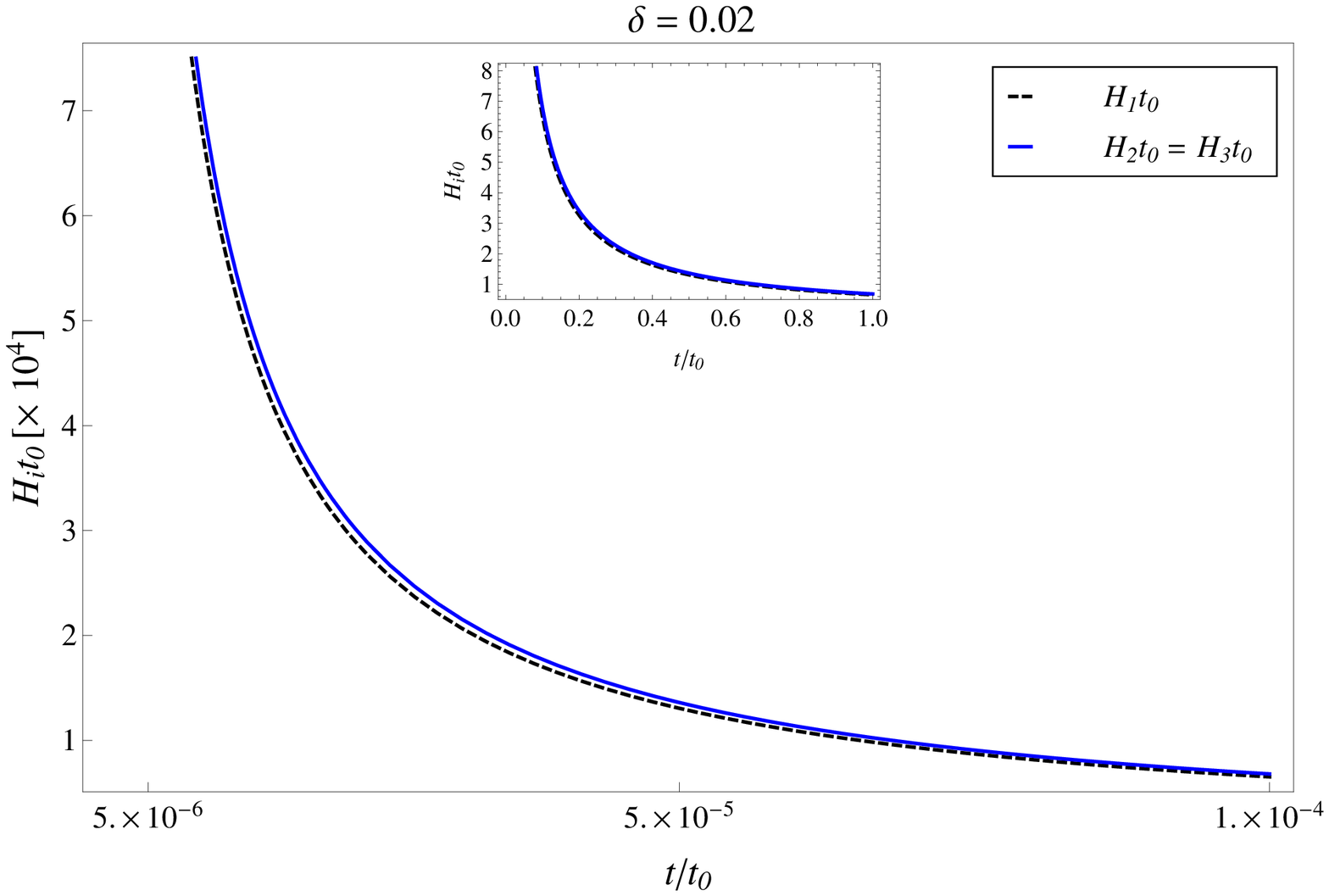}
\includegraphics[trim=0.55cm 0cm 2cm 0cm, clip=true,scale=0.44]{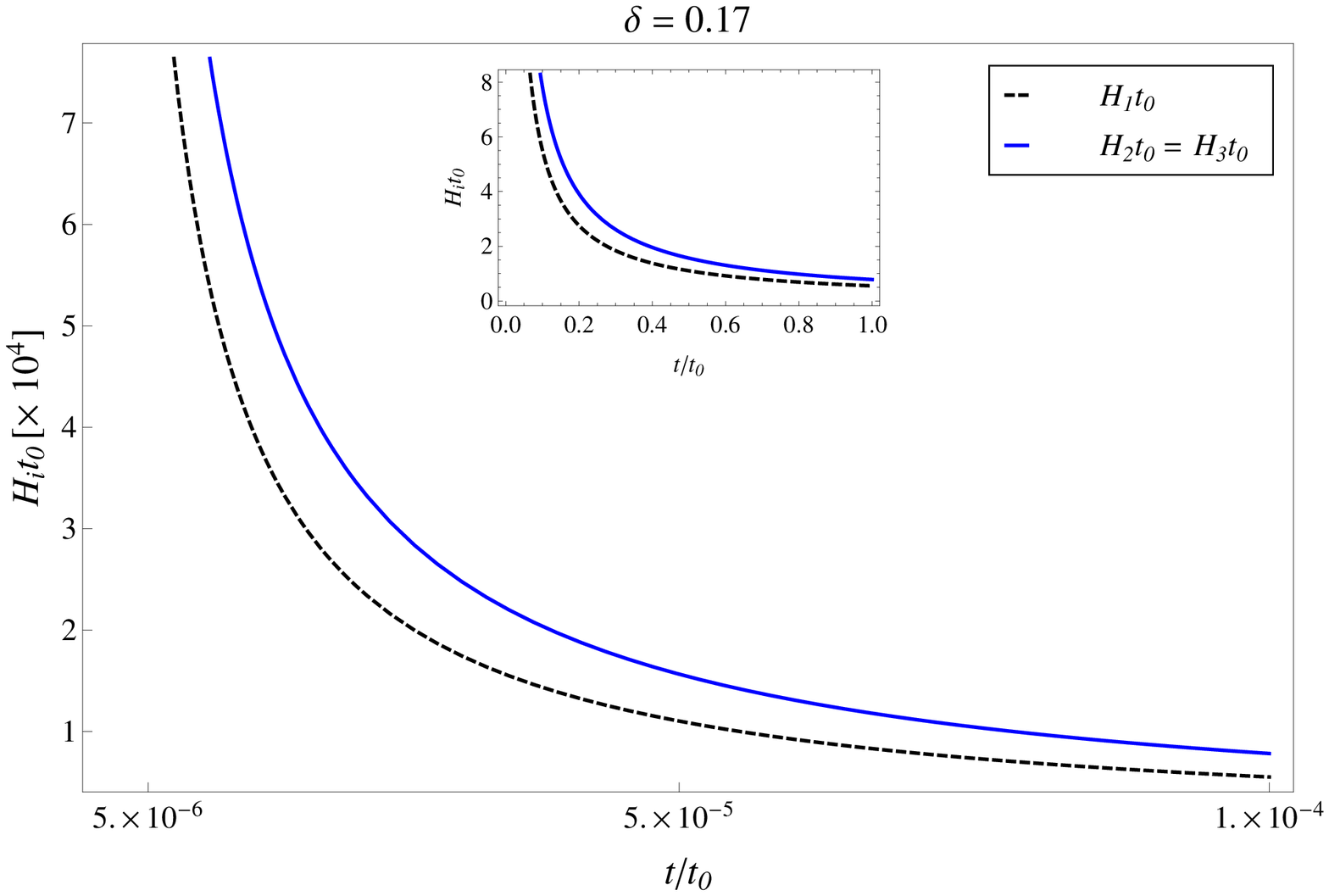}
\par\end{centering}
\caption{Directional Hubble parameters $H_i(t)$, for $i=1,2,3$. Here we adopt exaggerated values for $\delta$
in order to emphasize the effect of the deviation from the FLRW metric. Time goes from the 
decoupling ($t/t_0 \simeq 10^{-6}$) to the present time $t=t_0$ in the secondary graphic. As we can see, larger
values of $\delta$ increase differences among the directional Hubble parameters.}
\label{H_graphic}
\end{figure}

\begin{figure}
\begin{centering}
\includegraphics[trim=0.6cm 0cm 2.5cm 0cm, clip=true,scale=0.52]{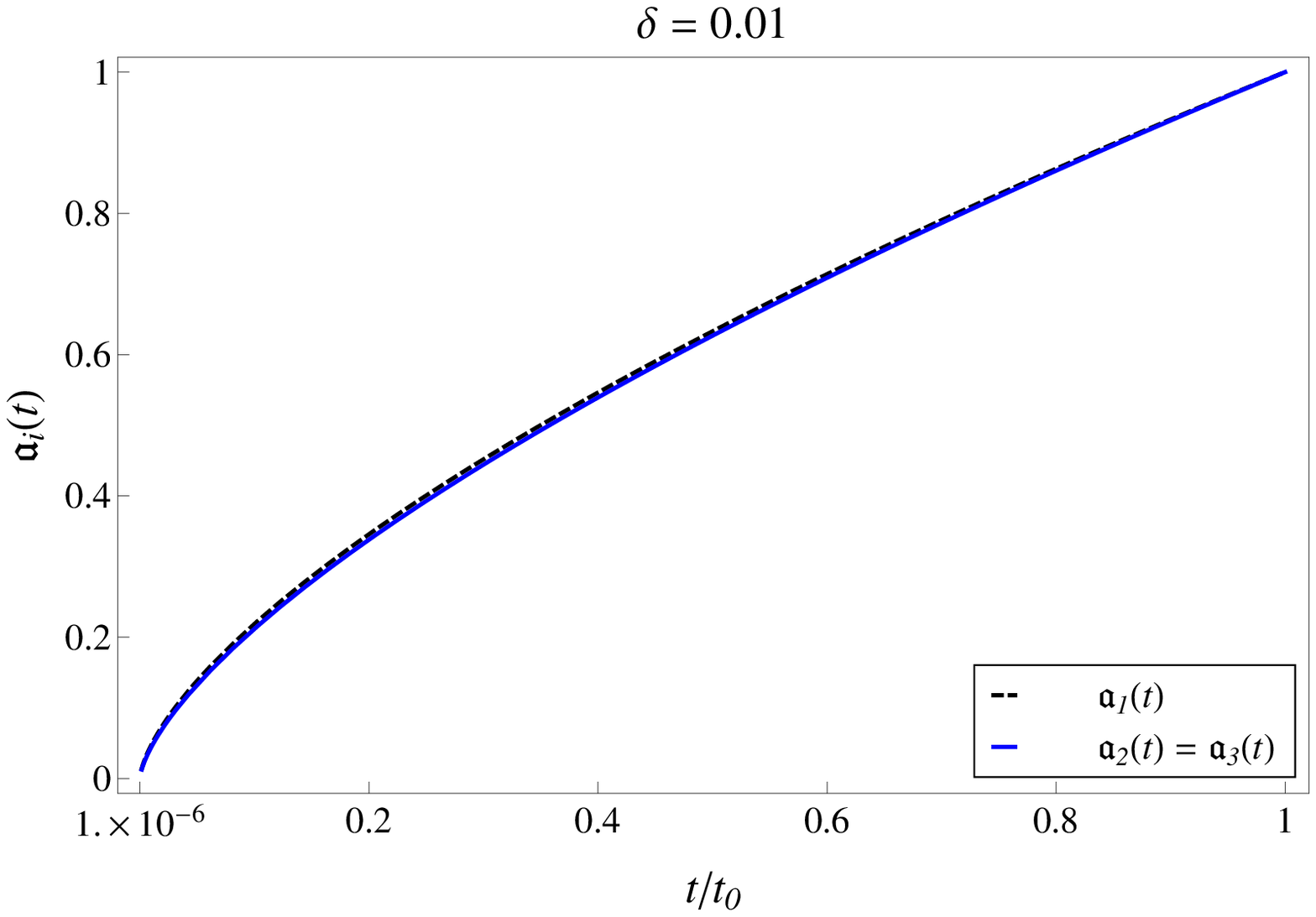}
\includegraphics[trim=0.3cm 0cm 2.5cm 0cm, clip=true,scale=0.52]{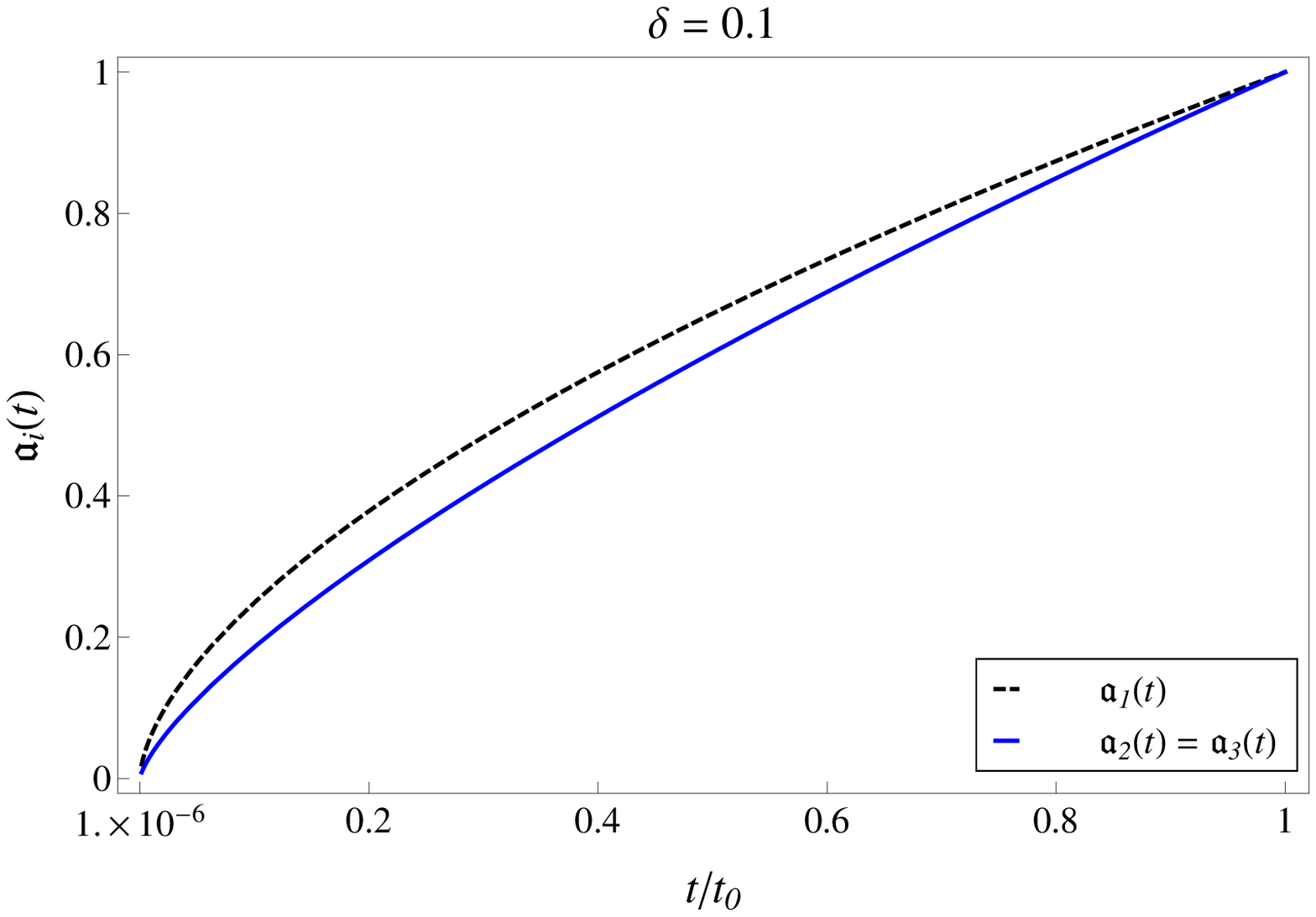}
\par\end{centering}
\caption{Normalized scale factors $\mathfrak{a}_i(t)$ and their dependence on $\delta$, the parameter
that deviates the solution (\ref{a1_normalized})-(\ref{a2_normalized}) from the FLRW spacetime. 
Here we adopt exaggerated values for $\delta$
in order to emphasize the effect of the deviation from the FLRW metric. Time goes from the 
decoupling ($t/t_0\simeq 10^{-6}$) to the present time $t=t_0$.}
\label{a_graphic}
\end{figure}

With the mean Hubble factor for a radiation plus matter universe, given by Eq. (\ref{Hubble_rm}), 
we are able to find out the factor (\ref{mu}) and solve the directional Hubble parameters (\ref{H1_solution})
 and (\ref{Hj_solution}). By doing that, we have
\begin{equation}
\mu (t)=4\Omega_{m,0}^3 \mathcal{V}_{rm},
\end{equation} 
and, consequently,
\begin{align}
H_1 t_0 = & \frac{2}{3} \left(1-\delta \right) \left( \frac{\mathcal{V}_m}{\mathcal{V}_{rm}}\right)\left( \frac{t_0}{t} \right), \label{H1_final}\\
H_2 t_0 =  H_3 t_0 = & \frac{2}{3} \left(1+\delta \right) \left( \frac{\mathcal{V}_m}{\mathcal{V}_{rm}}\right)\left( \frac{t_0}{t} \right).
\label{H2_final}
\end{align} 
With all directional Hubble factors available, the corresponding normalized scale factors are straightforwardly 
obtained:
\begin{align}
\mathfrak{a}_1 (t) = & \left(1-\frac{1-\left( \frac{t}{t_0}\right)^2}{1+5\frac{\mathcal{V}_{rme}}{\mathcal{V}_m}}  \right)^{\frac{1}{3}\left(1-\delta \right)},  \label{a1_normalized}  \\
\mathfrak{a}_2 (t) = \mathfrak{a}_3 (t) = & \left(1-\frac{1-\left( \frac{t}{t_0}\right)^2}{1+5\frac{\mathcal{V}_{rme}}{\mathcal{V}_m}}  \right)^{\frac{1}{3}\left(1+\delta \right)}.
\label{a2_normalized}
\end{align}
Assuming that $\delta\ll 1$, it is possible to express the normalized 
scale factors as $\mathfrak{a}_1 (t) \simeq \mbox{FLRW} + \delta_1$ and 
$\mathfrak{a}_2 (t) =\mathfrak{a}_3 (t) \simeq \mbox{FLRW} + \delta_2$, in which  $\delta_1$ 
and $\delta_2$ stand for small deviations from the FLRW metric. 
As we can see in Fig. \ref{a_graphic}, such  deviations firstly grow with time and secondly 
decreases in order to provide an almost FLRW metric.
Before the present day or $t=t_0$, the difference among scale factors can be as small as $\delta$ is.

 \section{Constraining the bumblebee field using cosmic anisotropies}
In this section, we  use recent values for the CMB multipoles provided by the Planck 
Collaboration \cite{Planck1,Planck2} in order to constrain the bumblebee field.  
 As we said, we assume that the Lorentz invariance or symmetry is spontaneously broken during the decoupling between 
 matter and radiation, producing then a small deviation from the FLRW metric.
 Our conventions and notation follow closely those ones of Ref. \cite{Russell}
in which the Bianchi I spacetime was studied in the general relativity realm.

 \subsection{Quantifying anisotropies}
We intend to produce a cosmological solution in which the universe turns into an isotropic world
at least as $t\rightarrow t_0$, in which $t_0$ indicates the present time or the age of the universe.
 That is, a universe described by Eqs. (\ref{H1_solution})-(\ref{Hj_solution})
  emerges from the decoupling anisotropically and
 becomes isotropic as $t\rightarrow t_0$. Our interest here is from the decoupling to the present time,
  where we had a matter-dominated universe. Like Russell \textit{et al.} \cite{Russell}, focusing on that period means that
  we can ignore the influence of the dark energy or, equivalently, put aside the cosmological
  constant in the modified field equations (\ref{modified}).
  
 According to Refs. \cite{Bronnikov,Saha}, a criterion to provide isotropization of the Bianchi 
 I model is given by the relation
 \begin{equation}
\lim_{t \rightarrow \infty} \frac{a_i}{a}= \mbox{constant} >0, \ \mbox{for}\  i=1,2,3,
\label{criterion}
\end{equation}   
with $a=(a_1a_2a_3)^{1/3}=\mathcal{V}^{1/3}$, being $\mathcal{V}$ known as the volume 
factor defined in Eq. (\ref{volume}) in terms
of the mean Hubble parameter. Spacetimes that satisfy (\ref{criterion}) become isotropic during the late-time 
expansion. That is, a metric like (\ref{Metric}) will turn into an isotropic metric if Eq. (\ref{criterion}) is true.  
In particular, for $\mbox{constant}=1$, we have
a FLRW metric in the present time. Such a criterion in Eq. (\ref{criterion}) is also called isotropization.
Like Ref. \cite{Russell}, we relax the isotropization condition  and assume 
that (\ref{criterion}) is valid as $t\rightarrow t_0$, producing
then an almost isotropic universe today. As we can see from Eqs. (\ref{a1_normalized})-(\ref{a2_normalized}),
 this criterion of isotropization for our 
solution is valid as $t \rightarrow t_0$. That is,
\begin{equation}
\lim_{t \rightarrow t_0} \frac{\mathfrak{a}_i}{\mathfrak{a}}= 1, \ \mbox{for}\  i=1,2,3,
\end{equation}
with $\mathfrak{a}=(\mathfrak{a}_1\mathfrak{a}_2\mathfrak{a}_3)^{1/3}=\mathcal{V}^{1/3}$. The isotropization occurs during a matter-dominated universe even after triggering the bumblebee field in the decoupling. 
A slight deviation from isotropy is view in Figs. \ref{H_graphic}-\ref{a_graphic}
 (deviation which depends on $\delta$) after the decoupling. Then the universe
converges to an almost FLRW universe.  

 Following Russell \textit{et al.} \cite{Russell}, who worked on the Einsteinian context, we adopt optical 
 scalars in order to quantify the anisotropic feature of the Bianchi I 
 metric in the bumblebee gravity context. Such scalars can be defined by 
 \begin{equation}
 A\equiv\frac{1}{3} \sum_{i=1}^{3}\left(\frac{H_i^2-H^2}{H^2} \right),
 \label{A}
 \end{equation}
 where $A$ is the mean anisotropy parameter, and
 \begin{equation}
 \sigma^{2}\equiv \sigma_{\mu\nu}\sigma^{\mu\nu}=\frac{3}{2} A H^2
 \label{shear}
 \end{equation}
 is the shear scalar defined in terms of the shear tensor $\sigma_{\mu\nu}$. The shear tensor is related to the 
 distortion of a spatial region, that is to say, an initially spherically symmetric configuration 
 of objects (for example, particles) turns into to an ellipsoidal shape due to the anisotropic
 feature of the spacetime, such that $\sigma^{2}$ 
 indicates the local distortion rate. It is important to note that both, $A$ and $\sigma^2$, should be
 small quantities at the present time. In particular, as we will point out, $A$ is given by the tiny parameter
 $\delta$ (the deviation from the standard metric), and $\sigma^2$ goes to zero as $t \rightarrow t_0$.

Another important scalar adopted here is the expansion scalar, associated with the expansion rate/Hubble parameter by
 \begin{equation}
 \Theta = 3H,
 \label{expansion}
 \end{equation}
 which, as we will see, alongside the shear scalar compounds an interesting relation to measure the amount
 of anisotropy in a given spacetime.  
 
In this article, the main relation in order to quantify anisotropies will be the ratio between shear and expansion
scalars. According to Maartens \textit{et al.} \cite{Maartens},  such a ratio can be directly related to the CMB data via
 \begin{equation}
 \frac{|\sigma_{\mu\nu}|}{\Theta}<\frac{5}{3}\epsilon_1+3\epsilon_2+\frac{3}{7}\epsilon_3,
 \label{Multipoles}
 \end{equation}
 with $|\sigma_{\mu\nu}|=(\sigma_{\mu\nu}\sigma^{\mu\nu})^{1/2}$ and $\epsilon_1$, $\epsilon_2$, and $\epsilon_3$ 
 indicating limits on the CMB dipole, quadrupole, and octopole,
 respectively. It is worth emphasizing that the relation (\ref{Multipoles}) is a model-independent approach 
 and, as we will see,  it will provide an upper bound on the magnitude of the bumblebee field.
 
 \begin{figure}
\begin{centering}
\includegraphics[trim=0.4cm 0cm 1.2cm 0cm, clip=true,scale=0.5]{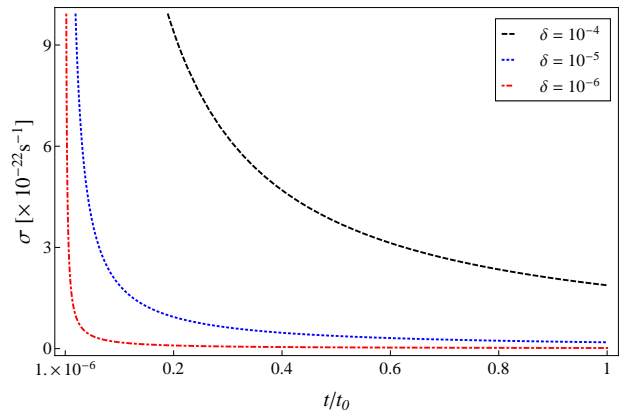}
\par\end{centering}
\caption{Shear scalar for some values of $\delta$, parameter which deviates the geometry from the FLRW.
Time goes from the decoupling ($t/t_0 \simeq 10^{-6}$)  to the present time $t=t_0$. The dependence of the
amount of shear on the parameter $\delta$ gets evident in this graphic.}
\label{shear_graphic}
\end{figure}

\subsection{Upper bound on the bumblebee field using the CMB data} 
With either the directional Hubble parameters or the scale factors, we can also quantify anisotropies and try to 
constrain the bumblebee field or the deviation from isotropy $\delta$ using the Planck data.
Before that, we are going to calculate, from the directional Hubble parameters (\ref{H1_final})-(\ref{H2_final}),
 the mean anisotropic parameter (\ref{A}). It follows that
\begin{equation}
A=\frac{8\delta^2}{\left(3+\delta \right)^2}.
\end{equation} 
As we can see, the mean anisotropy parameter $A$ is constant during the period studied here. However, its value is
very small as we are going to see after constraining $\delta$. On the other hand, the shear scalar is a time-dependent
parameter, whose final form, for the geometry studied here, reads
\begin{equation}
\sigma^2 = 3 \left(  \frac{H_0^2 \Omega_{m,0}t \delta}{\mathcal{V}_{rm}}  \right)^2.
\end{equation}
From Fig. \ref{shear_graphic}, one sees that the shear scalar decreases as $t$ increases. For the time present, its value is very small. 
Lastly, the expansion scalar (\ref{expansion}) is straightforwardly obtained from the directional Hubble parameters and the
mean Hubble parameter. It reads
\begin{equation}
\Theta=\frac{9}{2}\left(1+\frac{\delta}{3} \right) \frac{H_0^2 \Omega_{m,0}t}{\mathcal{V}_{rm}}.
\end{equation}
Accordingly,  $\Theta$ decreases with time as well. Hence the optical scalars are tiny quantities in the
present-day universe as we would expect it. With those scalars, the square ratio between the shear and
the expansion scalar reads simply
\begin{equation}
\left( \frac{\sigma}{\Theta}\right)^2 =\frac{4 \delta^2}{3 \left(3+\delta \right) ^2},
\label{ratio}
\end{equation}
which is constant (very small) with time. 

In order to constrain $\delta$ and, consequently, the bumblebee field, we follow Refs. \cite{Stoeger_a,Stoeger_b}
 in which the
Maartens \textit{et al.} \cite{Maartens} approach, mentioned in Eq. (\ref{Multipoles}), 
was adapted to the COBE data. Here, like Ref. \cite{Russell}, 
we apply the Planck data in that model-independent approach. Firstly, the ratio between
the shear and the expansion scalars should be given in terms of averages. The limits on the dipole $\epsilon_1$, 
quadrupole $\epsilon_2$ and octopole $\epsilon_3$ in Eq. (\ref{Multipoles}) are then written as averages
from the root-mean-square (rms) values of the multipole moments, namely\footnote{This is the corrected 
 form for $\left\langle \epsilon_l \right\rangle^2$ published in the erratum \cite{Stoeger_b}.}
\begin{equation}
\left\langle \epsilon_l \right\rangle^2=\frac{\left(2l+1 \right)\left(2l \right)!}{2^l \left(l! \right)^2}\left( \frac{\Delta T_l}{T_0} \right)^2,
\label{epsilons}
\end{equation}
where $\left\langle...\right\rangle$ means now average, $l$ stands for the multipole parameter 
($l=1$ for dipole, $l=2$ for quadrupole, and $l=3$ for octopole), and
$T_0=2.725$ K is the average temperature of the CMB radiation. According to Stoeger
\textit{et al.} \cite{Stoeger_a}, $\Delta T_2^2=Q^2_{rms}$ and $\Delta T_3^2=O^2_{rms}$ are, for example,
squares of the rms quadupole and octopole amplitudes, respectively. Also
$\Delta T_l^2$  are squares of the rotationally invariant rms  
multipole moments in the usual Legendre polynomial expansion of the two-point correlation function of 
the temperature anisotropy, they are written as
\begin{equation}
\Delta T_l^2 = \frac{1}{4\pi} \sum_{m=-l}^{+l} \vert a_{lm} \vert ^2,
\end{equation} 
in such a way that $a_{lm}$ are the coefficients of the expansion in spherical harmonics of the CMB difference of temperature in a given direction. However, as mentioned in Ref. \cite{Stoeger_a}, 
the CMB data, in general, is presented by means of 
 power spectrum graphics. Thus, we need to relate the averages of the $\epsilon$'s values
 with the CMB power spectrum. 
   Indeed, the coefficient $\Delta T_l^2$  is related to the power spectrum $\mathcal{D}_l$ of the CMB anisotropies  by
\begin{equation}
\Delta T_l^2 =  \frac{2l+1}{2l\left(l+1 \right)} \mathcal{D}_l.
\label{Delta_2}
\end{equation}
Therefore, substituting Eq. (\ref{Delta_2}) into Eq. (\ref{epsilons}), now we have  
a relation between averages of $\epsilon$'s and the power spectrum.

For our considerations, following Refs. \cite{Russell,Stoeger_a}, we turn off the dipole 
term $(\epsilon_1=0)$, for it is a term
whose origin is the solar system motion.
And according to the Planck Collaboration \cite{Planck1,Planck2}, the values of the power spectrum for the quadrupole 
and octopole moments are
\begin{equation}
\mathcal{D}_2 \simeq 300 \left[ \mu K^2 \right] \hspace{0.5cm} \mbox{and} \hspace{0.5cm} \mathcal{D}_3 \simeq 1000 \left[ \mu K^2 \right].
\end{equation}
Using that values in Eq.  (\ref{Delta_2}) and, consequently, in Eq. (\ref{epsilons}), one has the average for the coefficients of the formula (\ref{Multipoles}), namely 
\begin{equation}
\left\langle \epsilon_2 \right\rangle = 1.1 \times 10^{-5} \hspace{0.5cm} \mbox{and} \hspace{0.5cm} \left\langle \epsilon_3 \right\rangle= 2.6 \times 10^{-5}, 
\end{equation}
which give us an upper limit on the average of (\ref{Multipoles}), that is to say, 
\begin{equation}
\left\langle  \frac{\vert \sigma_{\mu\nu} \vert}{\Theta}  \right\rangle < 4.5 \times 10^{-5}. 
\end{equation}
With the aid of Eq. (\ref{ratio}), the above inequality then leads to an upper bound on the deviation from isotropy of 
the solution (\ref{H1_solution})-(\ref{Hj_solution}), given by the parameter $\delta$, i.e.,  
\begin{equation}
\delta < 10^{-4}
\label{delta_bound}
\end{equation}
is the upper limit on the amount of anisotropy. In order for the model presented here to 
be in agreement with the CMB data and the statement 
that our universe is almost isotropic, the deviation from the FLRW metric should be a tiny quantity as we read in Eq. (\ref{delta_bound}). 

With an upper bound on $\delta$, we are able to constrain the norm of the bumblebee field in the VEV,
due to the definition of the Lorentz-violating parameter $\ell = \xi b^2$.
It follows from (\ref{delta}) that
\begin{equation}
b^2 < 10^{39} \hspace{0.1cm} \mbox{kg}\ \mbox{m}\ \mbox{s}^{-2} \sim 10^{51}\hspace{0.1cm} \mbox{eV}^{2}.
\end{equation}
Alternately, adopting the best upper bound on the Lorentz-violating parameter to date (according to Ref. \cite{Guiomar},
 it is $\ell < 10^{-23}$), we are able to constrain, for the first time in the literature, 
 the coupling constant $\xi$. Thus, we have $\xi<10^{-62}\mbox{kg}^{-1}\mbox{m}^{-1}\mbox{s}^2 \sim 10^{-74}\hspace{0.1cm} \mbox{eV}^{-2}$ for that constant.

\section{Final remarks}
Using the Bianchi I geometry in a Lorentz symmetry breaking context gave us a source for
cosmological anisotropies. The model presented here is able to be conceived of as a small deviation
from the standard cosmology. It assumes an anisotropic description of  spacetime, 
by means of the Bianchi I metric instead of the FLRW metric, and a source for anisotropies, the bumblebee field.
With values for the CMB quadrupole and octopole moments, we showed that it is possible to think of the 
model presented in this article as a small deviation from the FLRW metric. 

The Lorentz violation is assumed to be triggered from the decoupling period between radiation and matter 
when the bumblebee field assumes the VEV.
Hence, the bumblebee field vector points toward a given direction, producing then different directional
Hubble parameters and a preferred axis. From the mentioned CMB multipoles, we constrained the bumblebee field
or its VEV and the coupling constant between that field and the geometry. As far as we know,
it is the very first attempt at constraining the bumblebee field using cosmological observations.

\section*{Acknowledgments}
RVM thanks Fundação Cearense de Apoio ao Desenvolvimento
Científico e Tecnológico (FUNCAP), Coordenação de Aperfeiçoamento de Pessoal de Nível Superior (CAPES), 
and Conselho Nacional de Desenvolvimento Científico e Tecnológico (CNPq, Grant no 307556/2018-2) 
for the financial support. We thank an anonymous referee for valuable suggestions.


\end{document}